\documentclass[prd, preprint, nofootinbib]{revtex4}

\usepackage{graphicx}
\usepackage{amsfonts}
\usepackage{latexsym}
\usepackage{amssymb}
\usepackage{epsfig}

\begin{document}

\title{ 
Gamma ray emission in Fermi bubbles \\
 and Higgs portal dark matter
}

\author{Nobuchika Okada}
 \email{okadan@ua.edu}
 \affiliation{
Department of Physics and Astronomy, 
University of Alabama, Tuscaloosa, Alabama 35487, USA
}

\author{Osamu Seto}
 \email{seto@phyics.umn.edu}
 \affiliation{
 Department of Life Science and Technology,
 Hokkai-Gakuen University,
 Sapporo 062-8605, Japan
}

%

\begin{abstract}
It has been recently pointed out that 
 the excess of the gamma ray spectrum in the Fermi bubbles 
 at low latitude can be well explained 
 by the annihilation of dark matter particles. 
The best-fit candidate corresponds to the annihilation 
 of a dark matter with mass of around $62$ GeV 
 into $b \bar{b}$ with the cross section, 
 $\sigma v \simeq 3.3 \times 10^{-26}$ cm$^3$/s,  
 or the annihilation 
 of a dark matter with mass of around $10$ GeV 
 into a tau lepton pair with the cross section, 
 $\sigma v \simeq 5.6 \times 10^{-27}$ cm$^3$/s. 
We point out that the Higgs portal dark matter models 
 are perfectly compatible with this interpretation 
 of the dark matter annihilation, 
 satisfying other phenomenological constraints. 
We also show that the parameter region which reproduces 
 the best-fit values can be partly explored 
 by the future direct dark matter search at the XENON1T.

\end{abstract}


\preprint{HGU-CAP 026} 

\vspace*{3cm}
\maketitle


\section{Introduction}

Recently, the gamma ray bubbles found in the Fermi-LAT data,
 the so-called Fermi bubbles~\cite{Su:2010qj},
 have received a fair amount of attention, 
 and their spectrum has been intensively studied. 
It has been pointed out in Ref.~\cite{Hooper:2013rwa}
 that the gamma ray spectrum at the low latitude region 
 shows an extra contribution in the energy range of $E \sim 1 - 4$ GeV, 
 while the spectrum at high latitude region can be 
 reasonably explained by the inverse Compton scattering. 
It has been shown in Ref.~\cite{Hooper:2013rwa} that 
 the excess can originate from the annihilation 
 of dark matter particles: 
 a $10$ GeV dark matter annihilating into a tau lepton pair 
 with the cross section (times relative velocity)
 $\sigma v = 2 \times 10^{-27}\, {\rm cm^3/s}$
 or a $50$ GeV dark matter annihilating into quarks 
 with the cross section $\sigma v = 8 \times 10^{-27}\, {\rm cm^3/s}$. 
Similarly and more recently, the authors of Ref.~\cite{Huang:2013pda} have claimed 
 that the excess is best fit by 
 a $10$ GeV dark matter annihilating into a pair of tau leptons 
 with $\sigma v \simeq 5.6 \times 10^{-27}\, {\rm cm^3/s}$
 or a $62$ GeV dark matter annihilating into $b \bar{b}$
 with $\sigma v \simeq 3.3 \times 10^{-26}\, {\rm cm^3/s}$. 
Interestingly, the magnitude of annihilation cross section
 favored by these analyses is close to 
 the typical thermal annihilation cross section, 
 $\sigma v \simeq 3 \times 10^{-26}\, {\rm cm^3/s}$, 
 for a weakly interacting massive particle dark matter 
 to reproduce its correct thermal relic abundance 
 of $\Omega_{\rm DM} h^2 \simeq 0.1$.

Besides the Fermi bubbles, the data of gamma rays from subhalos 
 also show a similar spectrum shape consistent 
 with the dark matter annihilation scenario: 
 $8-10$ GeV dark matter annihilating to tau leptons 
 with $\sigma v \simeq (1-2) \times 10^{-27}\, {\rm cm^3/s}$
 or $30-60$ GeV dark matter annihilating to $b \bar{b}$ with 
 $\sigma v \simeq (5-10) \times 10^{-27}\, {\rm cm^3/s}$~\cite{Berlin:2013dva}. 
See also \cite{Goodenough:2009gk, Hooper:2010mq} 
 for similar discussions.

In this paper, we point out that 
 the so-called Higgs portal dark matter scenario 
 suits the interpretation of the dark matter annihilation 
 for the gamma ray spectrum from the Fermi bubbles. 
We consider two simple Higgs portal dark matter models 
 with a real scalar dark matter being singlet 
 under the Standard Model (SM) gauge groups. 
The first model is one of the simplest extensions of  
 the SM and we introduce the SM gauge singlet real scalar 
 along with a $Z_2$ parity 
 (for an incomplete list, see,  
 e.g.,~\cite{McDonald:1993ex,Burgess:2000yq, 
 Davoudiasl:2004be,Kikuchi:2007az,KMNO}). 
The scalar dark matter with mass of around 60 GeV 
 mainly annihilates into $b\bar{b}$ 
 through the SM Higgs boson in the $s$ channel. 
The other model is the Higgs portal dark matter 
 realized in the two-Higgs-doublet extension of the SM. 
In this model, the scalar dark matter with mass of around 10 GeV 
 mainly annihilates into a tau lepton pair 
 through the Higgs bosons exchange in the $s$ channel. 
We show that both of the models can account 
 for the excess of the gamma ray spectrum from the Fermi bubbles, 
 satisfying the cosmological condition for the observed relic abundance 
 as well as the constraint from the current direct dark matter 
 search experiments. 
See Ref.~\cite{Hagiwara:2013qya} 
 for a supersymmetric model with a 10 GeV neutralino dark matter 
 which can account for the Fermi bubble excess 
 through the neutralino pair annihilation to tau leptons 
 mediated by light scalar tau leptons. (See also, e.g., Ref~\cite{Kyae:2013qna}.)

\section{Standard model Higgs portal scalar dark matter} 

At first, we show that a gauge singlet scalar dark matter $\phi$
 with the mass about $60$ GeV has the desired property 
 to account for the Fermi bubble excess. 
We only add a real scalar $\phi$ to the SM particle contents 
 along with a $Z_2$ parity, under which 
 the scalar is odd while the SM particles are even. 
The Lagrangian is given by
\begin{equation}
 {\cal L} = {\cal L}_{\rm SM} + 
 \frac{1}{2}(\partial\phi)^2 - \frac{1}{2} M_{\phi}^2 \phi^2
 - \frac{1}{2} c |\Phi|^2\phi^2 -\lambda_\phi \phi^4,
\end{equation}
 where $\Phi$ denotes the SM Higgs doublet field, and 
 $c$ is a dimensionless coupling constant. 
After the electroweak symmetry breaking, 
 the dark matter mass is given by 
 $m_\phi^2= M_\phi^2 + c v^2/2$ 
 with the Higgs vacuum expectation value ($v$), 
 and the interaction terms between the scalar dark matter 
 and the physical Higgs boson ($h$) are given by 
\begin{eqnarray}
 {\cal L}_{\rm int}= -\frac{c}{2} v h \phi^2 -\frac{c}{4} h^2 \phi^2.  
\end{eqnarray}
When $m_\phi < m_h/2$, the Higgs boson decays 
 invisibly, $h\rightarrow \phi \phi$, through the coupling. 
 From the LHC data, the branching ratio of 
 the invisibly decaying Higgs boson
 is constrained (at 3$\sigma$) as
%
${\rm BR}(h \rightarrow \phi \phi) \lesssim 0.35$~\cite{Hinv}, 
%
which leads to an upper bound on the coupling constant $c$. 

\begin{figure}[h,t]
\begin{center}
\epsfig{file=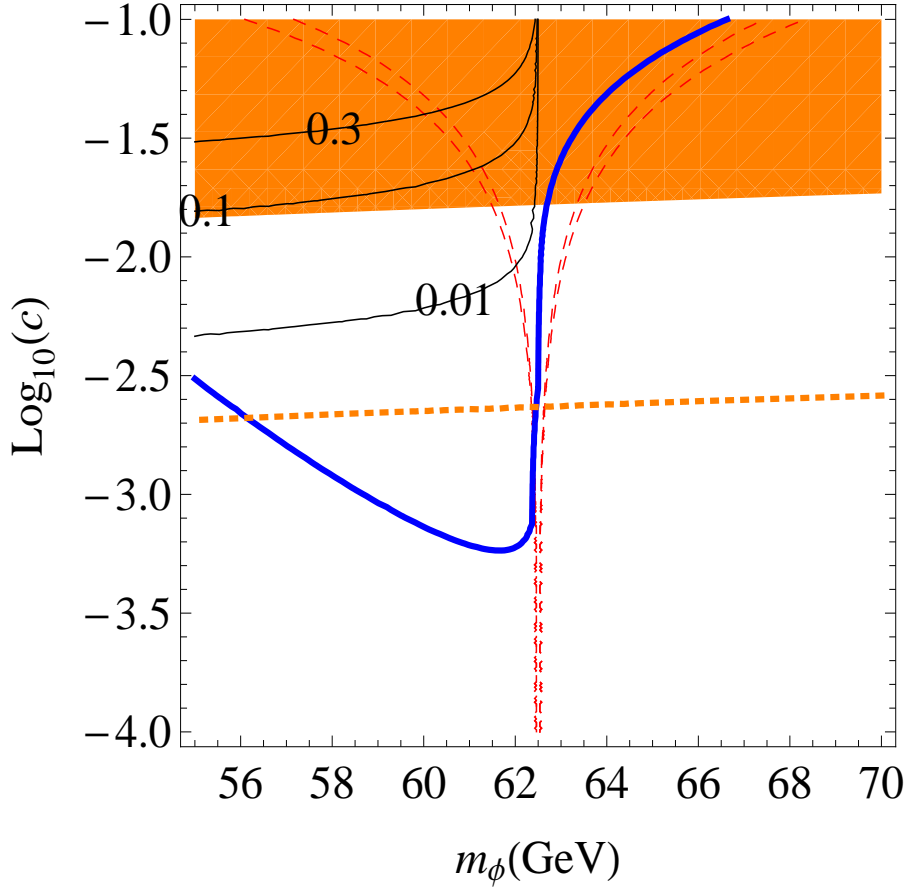, width=12cm, height=12cm, angle=0}
\end{center}
\caption{
Various constraints and predictions:
 $\Omega h^2 = 0.1$ (thick blue line),
 $2.81 \times 10^{-26} \leq (\sigma v)_0 \leq 3.99 \times 10^{-26}$ cm$^3$/s 
 (dashed red lines) claimed in Ref.~\cite{Huang:2013pda}, 
 and the contours corresponding to 
 ${\rm BR}(h \to \phi \phi)=0.01$, $0.1$, and $0.3$ (solid gray line). 
The shaded region is excluded by the direct dark matter search
 at XENON100(2012)~\cite{XENON100:2012}, 
 and the future expected sensitivity $4\times 10^{-47}\, {\rm cm}^2$ by 
 the XENON1T experiment~\cite{XENON1T} 
 is depicted by the horizontal dotted line. 
Here, we have fixed $m_h =125$ GeV.
 }
\label{Fig:bbPortal}
\end{figure}

The thermal relic abundance of the dark matter 
 is evaluated by solving the Boltzmann equation 
 for the number density of $\phi$:
\begin{equation}
 \frac{d n }{dt}+3H n =-\langle\sigma v\rangle( n^2 - n_{\rm EQ}^2),
\end{equation}
 with $H$ and $n_{\rm EQ}$ being the Hubble parameter and
 the dark matter number density in thermal equilibrium, 
 respectively~\cite{KolbTurner}. 
With a good accuracy, the resultant thermal relic abundance 
 is expressed as 
\begin{equation}
\Omega_{\rm DM} h^2 =
 \frac{1.1 \times 10^9 x_d \; {\rm GeV^{-1}}}
 {\sqrt{g_*}M_P\langle\sigma v\rangle}, 
\end{equation}
 where $M_P=1.22 \times 10^{19}$ GeV is the Planck mass, 
 $\langle\sigma v\rangle$ is the thermal averaged product 
 of the annihilation cross section and the relative velocity, 
 $g_*$ is the total number of relativistic degrees of freedom 
 in the thermal bath,  
 and $x_d = m_\phi/T_d$ with the decoupling temperature $T_d$. 
For the dark matter mass of around 60 GeV and 
 the Higgs boson mass of around 125 GeV,   
 the scalar dark matter dominantly annihilates 
 to the $b \bar{b}$ final state  
 through the Higgs boson exchange in the $s$ channel.

For a given dark matter mass, we identify the value of  
 the coupling constant ($c$) so as to reproduce 
 the observed relic abundance 
 $\Omega h^2 \simeq 0.1$~\cite{WMAP, Planck}. 
Note that for a fixed parameter set, the thermal averaged 
 cross section $\langle\sigma v\rangle$ determined 
 by the condition of $\Omega h^2 \simeq 0.1$ 
 is in general different from the present annihilation 
 cross section $(\sigma v)_0$ of the dark matter 
 relevant to the indirect search for dark matter. 
Here $(\sigma v)_0$ is simply given by the limit of 
  the vanishing relative velocity $v \rightarrow 0$, 
 rather than taking the thermal average. 
This difference is noteworthy for the dark matter mass 
 being close to the Higgs resonance pole 
 in its annihilation process, $m_\phi \simeq m_h/2$.

In Fig.~\ref{Fig:bbPortal}, we show the thick blue line  
 along which $\Omega h^2 = 0.1$ is satisfied. 
The region inside the dashed red lines corresponds 
 to the dark matter annihilation cross section 
 to the $b \bar{b}$ final state at the present Universe 
 in the range of $2.81 \times 10^{-26} \leq (\sigma v)_0 \leq 3.99 \times 10^{-26}$ cm$^3$/s. 
Recently it has been pointed out in Ref.~\cite{Huang:2013pda} that 
 this range of the annihilation cross section 
 gives the best fit for the gamma ray spectrum 
 from the Fermi bubbles for the dark matter mass 
 in the range of 56.9 $\leq m_{DM} \leq 68.7$ GeV. 
We have found in Fig.~\ref{Fig:bbPortal} that 
 the best fit for the gamma ray spectrum from the Fermi bubbles 
 and the observed relic abundance are simultaneously 
 realized by $c \simeq 10^{-3}$ and $m_\phi \simeq 62.5$ GeV. 
In the figure, the contours for the branching ratio 
 of the Higgs invisible decay 
 [${\rm BR} (h \to \phi \phi) = 0.01, 0.1$, and $0.3$] 
 are also shown. 
The shaded region is excluded by the null result of the direct 
 dark matter search at the XENON100~\cite{XENON100:2012} 
 while the horizontal dotted line denotes 
 the future reach by the XENON1T experiment~\cite{XENON1T}.

\section{Two-Higgs-doublet portal scalar dark matter}

Next let us consider the Higgs portal dark matter 
 realized in the two-Higgs-doublet extension of the SM, 
 namely, the so-called type-X two-Higgs-doublet model (THDM).
This type of THDM has been extensively studied  
 from the viewpoint of, especially, 
 nonvanishing neutrino mass~\cite{Aoki:2008av},
 and the results from dark matter direct~\cite{Aoki:2009pf}
 or indirect~\cite{Goh:2009wg} searches. 
In this model, a scalar dark matter with mass of 
 around $10$ GeV annihilates mainly into a tau lepton pair.

In the type-X model, the Yukawa interaction is given by 
\begin{eqnarray}
 {\cal L}_Y 
   = - y_{\ell_i}  \overline{L}^i \Phi_1 \ell_R^i
     - y_{u_i}  \overline{Q}^i \tilde{\Phi}_2 u_R^i
     - y_{d_i}  \overline{Q}^i \Phi_2 d_R^i + {\rm h.c.}, \label{eq:yukawa1}
\end{eqnarray}
 where $Q^i$ ($L^i$) is the ordinary left-handed quark (lepton) 
 in the $i$th generation, and $u_R^i$ and $d_R^i$ ($e_R^i$) are 
 the right-handed SU(2) singlet up- and down-type quarks 
 (charged leptons), respectively.   
Here, we have neglected the flavor mixing, for simplicity. 
The scalar potential for the two-Higgs doublets ($\Phi_1$ and $\Phi_2$) 
 is given by
\begin{eqnarray}
 V &=& -\mu_1^2 |\Phi_1|^2 -\mu_2^2 |\Phi_2|^2 - (\mu_{12}^2
  \Phi_1^\dagger \Phi_2 + {\rm h.c.}) \nonumber \\
 &&+ \lambda_1|\Phi_1|^4   +
  \lambda_2|\Phi_2|^4 + \lambda_3|\Phi_1|^2|\Phi_2|^2 +\lambda_4 |\Phi_1^\dagger \Phi_2|^2 \  +
   \left\{ \frac{\lambda_5}{2} (\Phi_1^\dagger
    \Phi_2)^2 + {\rm h.c.} \right\} \nonumber\\ 
 &&  + \frac{1}{2}\mu_\phi^2 \phi^2 +
  \lambda_\eta \phi^4 
   + ( \sigma_1 |\Phi_1|^2 + \sigma_2 |\Phi_2|^2 ) \frac{\phi^2}{2}.
\end{eqnarray}

Electric charge neutral components of the two-Higgs doublets 
 develop the vacuum expectation values as
\begin{eqnarray}
  \Phi_1
   = 
  \left( \begin{array}{c}
          0 \\
          \frac{v_1 + h_1}{\sqrt{2}} \\
         \end{array}
  \right), \hspace{1cm}
%
  \Phi_2
   = 
  \left( \begin{array}{c}
          0 \\
          \frac{v_2 + h_2}{\sqrt{2}} \\
         \end{array}
  \right),
\end{eqnarray}
 where $v^2=v_1^2+v_2^2 = (246~\rm{ GeV})^2$, 
 and we introduce the usual parametrization, 
 $\tan\beta=v_2/v_1$. 
The physical states ($h_1$ and $h_2$) are diagonalized 
 to the mass eigenstates ($h$ and $H$) as 
 \begin{eqnarray}
  \left( \begin{array}{c}
          h_1  \\
          h_2  \\
         \end{array}\right)
   = 
  \left( \begin{array}{cc}
          \cos\alpha & - \sin\alpha\\
          \sin\alpha & \cos\alpha\\
         \end{array}\right)
  \left( \begin{array}{c}
          H \\
          h \\
         \end{array}\right) .
 \end{eqnarray}
When the mixing angle $\alpha$ satisfies 
 the condition $\sin(\beta-\alpha) \simeq 1$, 
 the mass eigenstate $h$ is the SM-like Higgs boson.

In terms of the mass eigenstates, the (3-point) interactions 
 of the scalar dark matter with the Higgs bosons are given by 
\begin{eqnarray}
%
{\cal L}_\sigma \supset 
 - \frac{\sigma_1 \cos\alpha\cos\beta+\sigma_2\sin\alpha\sin\beta}{2}
  v H \phi^2 
- \frac{- \sigma_1 \sin\alpha\cos\beta+\sigma_2\cos\alpha\sin\beta}{2}
  v h \phi^2.
\label{HiggsInt}
 \end{eqnarray}
The Yukawa interactions with quarks and leptons 
 in Eq.~(\ref{eq:yukawa1}) can then be written as
\begin{eqnarray}
 {\cal L}_Y^{\rm Quarks} &\supset&  - \frac{m_{u^i} \sin\alpha}{v \sin\beta} H \bar u^i
  u^i - \frac{m_{u^i} \cos\alpha}{v \sin\beta} h \bar u^i u^i
  - \frac{m_{d^i}\sin\alpha}{v\sin\beta} H \bar d^i d^i 
  - \frac{m_{d^i} \cos\alpha}{v\sin\beta} h \bar d^i d^i , \\
{\cal L}_Y^{\rm Leptons} &\supset &
    -\frac{m_{\ell^i}}{v} \frac{\cos\alpha}{\cos\beta} H
    \bar \ell^i \ell^i +    \frac{m_{\ell^i}}{v} \frac{\sin\alpha}{\cos\beta} h
    \bar \ell^i \ell^i.
 \end{eqnarray}
In the following analysis, we fix the mixing angle 
 to give $\sin(\beta-\alpha)= 1$. 
Then, the coupling between the non-SM-like Higgs ($H$) 
 and the lepton is enhanced for $\tan\beta > 1$, 
 while the SM couplings between the Higgs boson  
 and the quarks remain the same as the SM ones. 
 From now on, we take $\tan\beta=5$ as a reference value.
In addition, we fix model parameters to make 
 the charged and $CP$-odd Higgs bosons heavy enough 
 to be consistent with the current experimental lower bound 
 and not to be involved in our analysis of the dark matter.~\footnote{
In the following, we will find the results 
 that the non-SM-like Higgs boson mass $\lesssim 30$ GeV, 
 and hence the mass splitting between this Higgs boson 
 and the charged and $CP$-odd Higgs bosons is large. 
We can check that even with the large mass splitting, 
 our model is consistent with the electroweak precision tests 
 when the charged and $CP$-odd Higgs bosons are well degenerate. 
Our model has enough freedom of free parameters 
 to realize such a mass spectrum, 
 keeping our results for dark matter physics intact. 
}

\begin{figure}[h,t]
\begin{center}
\epsfig{file=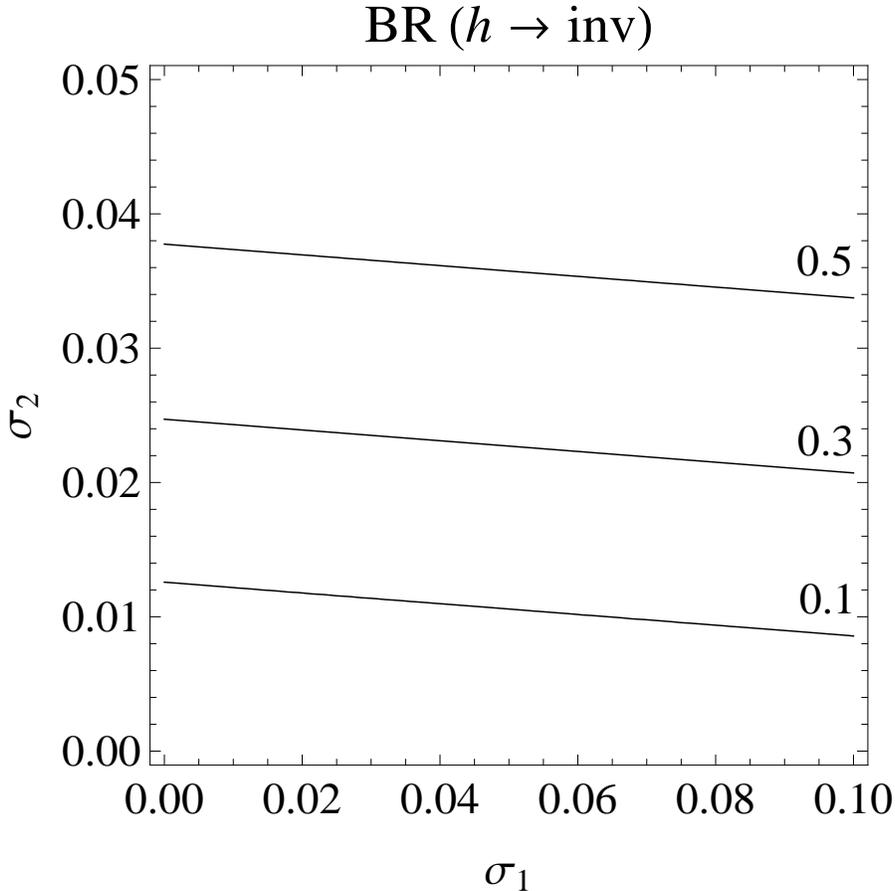, width=12cm,height=12cm,angle=0}
\end{center}
\caption{
 Contours of the invisible decay branching ratio 
 of the SM-like Higgs boson, 
 ${\rm BR}(h \rightarrow \phi \phi)=0.1$, $0.3$, and $0.5$, 
 respectively. 
We have taken $\tan \beta =5$, $\sin(\beta -\alpha)=1$, 
 and $m_\phi=10$ GeV. 
 }
\label{Fig:invisible}
\end{figure}

We first calculate the invisible decay width of 
 the SM-like Higgs boson into a pair of the scalar dark matters 
 through the interactions in Eq.~(\ref{HiggsInt}).~\footnote{
As we will see in the following, 
 the Higgs boson $H$ is light and the SM-like Higgs boson 
 also decays to a pair of the $H$ bosons. 
Since many free parameters are involved 
 in the decay process (see the Appendix), 
 we simply assume a negligible partial decay width
 for it in this paper. 
} 
The branching ratio of the invisible decay 
 ${\rm BR}(h \rightarrow \phi \phi)$ 
 is shown in Fig.~\ref{Fig:invisible}. 
We have found that the bound from the LHC data, 
 ${\rm BR}(h \rightarrow \phi \phi) \lesssim 0.35$~\cite{Hinv}, 
 is satisfied for $\sigma_2 \lesssim 0.02$, 
 almost independently of $\sigma_1$. 

\begin{figure}[h,b]
\begin{center}
\epsfig{file=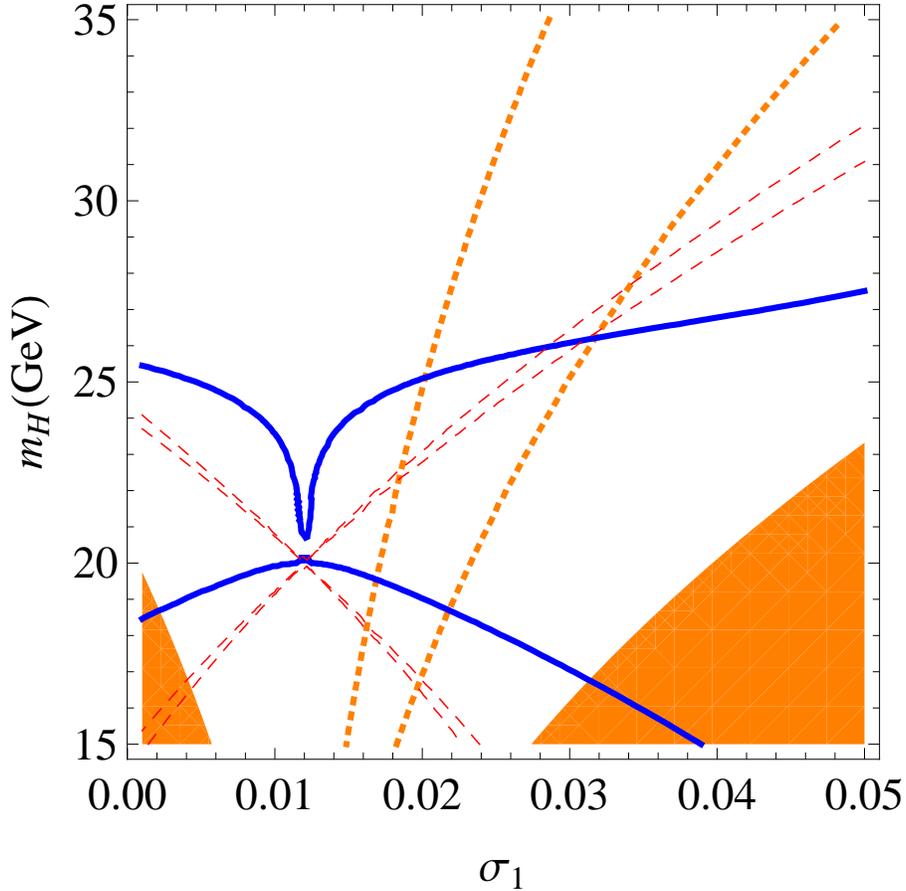, width=12cm,height=12cm,angle=0}
\end{center}
\caption{
Contours for $\Omega h^2 = 0.1$ (thick blue line) 
 and $2.81 \times 10^{-26} \leq (\sigma v)_0 \leq 3.99 \times 10^{-26}$ cm$^3$/s 
 (dashed red lines) claimed in Ref.~\cite{Huang:2013pda}. 
The shaded regions are excluded by the direct dark matter search 
 by the XENON100(2012) experiment~\cite{XENON100:2012}, 
 and the expected future sensitivity  
 $5\times 10^{-45}\, {\rm cm}^2$
 by the XENON1T experiment~\cite{XENON1T} 
 are depicted as the dotted lines. 
In this  analysis, we have fixed $\sigma_2=0.012$ 
 and $m_\phi=10$ GeV. 
}
\label{Fig:tautau012}
\end{figure}

Now we calculate the annihilation cross section 
 of the scalar dark matter dominated by 
 the $s$-channel Higgs bosons ($h$ and $H$) exchange.  
We evaluate the cross section as a function of 
 the coupling $\sigma_1$ and 
 the non-SM-like Higgs boson mass $m_H$
 with a fixed value for $\sigma_2 < 0.02$. 
For the dark matter with $m_\phi=10$ GeV,
 the annihilation mode into a tau lepton pair 
 through the $H$-boson exchange dominates 
 for suitable values of $\sigma_1$ and $m_H$.

\begin{figure}[h,b]
\begin{center}
\epsfig{file=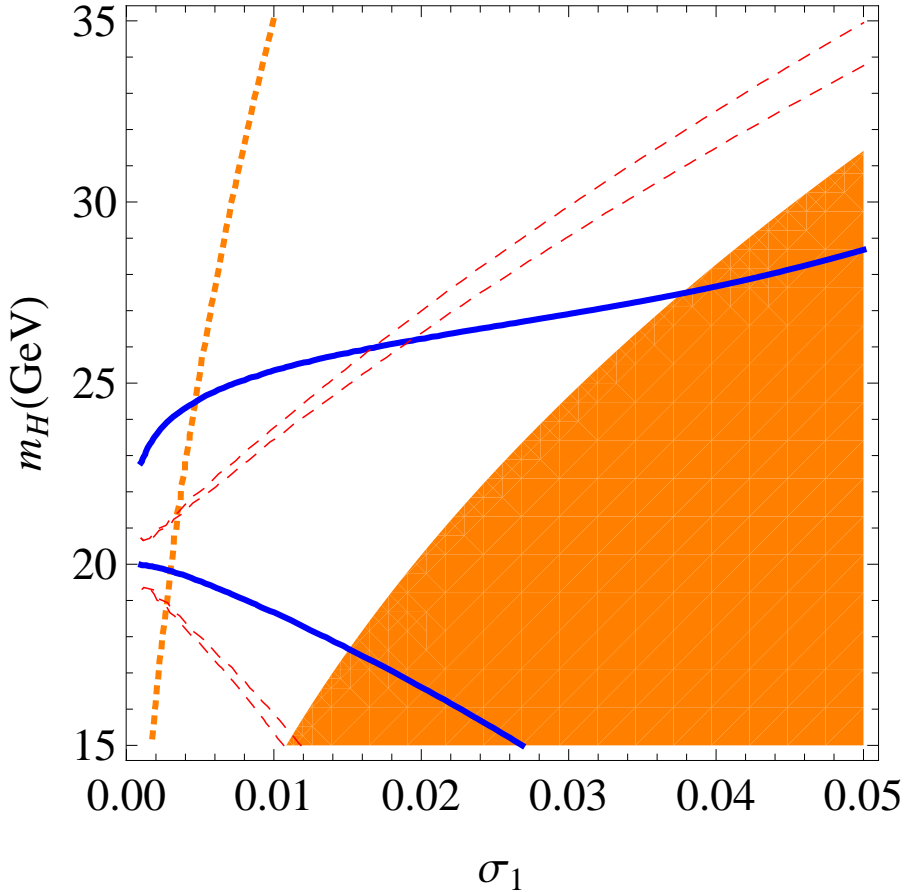, width=12cm,height=12cm,angle=0}
\end{center}
\caption{
Same as Fig.~\ref{Fig:tautau012} but for $\sigma_2=0$. 
 }
\label{Fig:tautau00}
\end{figure}

Figure~\ref{Fig:tautau012} shows the results  for $\sigma_2=0.012$. 
The thick blue line corresponds to the parameter set 
 which reproduces the thermal relic abundance 
 of the scalar dark matter $\Omega h^2=0.1$, 
 while the parameters between the two dashed red lines 
 provide the annihilation cross section, 
 $2.81 \times 10^{-26} \leq (\sigma v)_0 \leq 3.99 \times 10^{-26}$ cm$^3$/s~\cite{Huang:2013pda}. 
The shaded regions are excluded by the direct dark matter search
 at the XENON100(2012)~\cite{XENON100:2012}, 
 and the expected sensitivity by the XENON1T experiment~\cite{XENON1T} 
 is depicted by two dotted lines. 
We can see that near the resonance pole $m_{\phi} = m_H/2$, 
 the conditions for the thermal relic abundance 
 and the best-fit annihilation cross section 
 into a tau pair claimed in Ref.~\cite{Huang:2013pda} 
 are simultaneously satisfied for 
\begin{eqnarray}
 \sigma_1 \simeq 0.03,~~~~
 m_H  \simeq  26 \; {\rm GeV}.  
\end{eqnarray}
This region is found to be close to the sensitivity 
 of the direct dark matter search expected by the XENON1T experiment. 
Results for the same analysis but with $\sigma_2=0$ 
 are depicted in Fig.~\ref{Fig:tautau00}. 
In this case, we have found the solution for  
\begin{eqnarray}
 \sigma_1 \simeq 0.018,~~~~
 m_H  \simeq  26 \; {\rm GeV}.  
\end{eqnarray}
Interestingly, this parameter region can be tested 
 by the XENON1T experiment in the future. 

%
%

\section{Summary}
We have shown that a Higgs portal dark matter particle 
 annihilating into $b\bar{b}$ or $\tau^{+} \tau^{-}$ 
 through the $s$-channel exchange of Higgs boson(s) 
 very well suits the dark matter interpretation 
 in explaining the excess of the gamma ray spectrum 
 from the Fermi bubbles at low latitude, observed by the Fermi-LAT. 
In the simplest Higgs portal dark matter model 
 (``SM plus $\phi$'' model), 
 we have identified a model-parameter region 
 ($c \simeq 10^{-3}$ and $m_\phi \simeq 62.5$ GeV) 
 which can simultaneously satisfy the correct thermal relic abundance 
 and the best-fit value of the dark matter annihilation 
 cross section to explain the gamma ray excess~\cite{Huang:2013pda}. 
Very interestingly, the mass we have found 
 is almost the best-fit value claimed in Ref.~\cite{Huang:2013pda}. 
In our analysis, we see that the parameter region appears 
 near the SM Higgs resonance point ($m_\phi \sim m_h/2$),  
 and therefore a suitable dark matter mass is almost 
 fixed by the the SM Higgs boson mass. 
The SM Higgs boson is finally discovered with a mass of 
 around $m_h=125-126$ GeV. 
It is another interesting point that 
 in the simplest Higgs portal dark matter model, 
 the observed Higgs boson mass is compatible 
 with the dark matter interpretation 
 for the gamma ray excess in the Fermi bubbles. 
We have also considered the Higgs portal dark matter model 
 realized in the two-Higgs-doublet extension of the SM 
 (``type-X THDM plus $\phi$'' model). 
In this case, a scalar dark matter with $m_\phi = 10$ GeV 
 dominantly annihilates into a pair of tau leptons. 
We have identified a parameter region 
 which reproduces the best-fit region 
 corresponding to the dark matter annihilation into 
 a tau lepton pair~\cite{Huang:2013pda}, 
 as well as the observed thermal relic abundance. 
We have found that the parameter region is partly covered 
 by the expected sensitivity of the direct dark matter search 
 at the XENON1T.

Finally, analysis in Refs.~\cite{Hooper:2013rwa, Huang:2013pda} 
 has been done by assuming a 100\% annihilation fraction 
 for a selected annihilation mode. 
However, for a given concrete particle model, 
 there are various annihilation modes in general. 
It should be worth performing more detailed analysis  
 for the gamma ray spectrum based  on a concrete model 
 with a realistic annihilation fraction to various final states. 
The Higgs portal dark matter scenario presented in this paper 
 can be a good benchmark for the analysis.


\section*{Acknowledgments}
This work is supported in part by 
 the DOE Grant No. DE-FG02-10ER41714 (N.O).
We would like to thank Koji Tsumura for valuable comments.
%


\appendix

\section{Type-X two-Higgs-doublet-model}

\subsection{Decay width of Higgs bosons}

\subsubsection{Invisible decay width}
\begin{eqnarray}
  \Gamma_h^{(inv)} &=& (- \sigma_1 \sin\alpha\cos\beta+\sigma_2\cos\alpha\sin\beta)^2
          \frac{v^2}{32\pi m_h}\sqrt{1-\frac{4 m_\phi^2}{m_h^2}} , \\
  \Gamma_H^{(inv)} &=& (\sigma_1 \cos\alpha\cos\beta+\sigma_2\sin\alpha\sin\beta)^2
          \frac{v^2}{32\pi m_H}\sqrt{1-\frac{4 m_\phi^2}{m_H^2}} .
\end{eqnarray}

\subsubsection{Total decay width}
\begin{eqnarray}
  \Gamma_h &\simeq& \sin^2(\beta-\alpha)\Gamma(h_{SM} \rightarrow VV)
              + \left(\frac{\cos\alpha}{\sin\beta}\right)^2 \Gamma(h_{SM} \rightarrow q\bar{q}) \nonumber\\
          && + \left(\frac{\sin\alpha}{\cos\beta}\right)^2 \Gamma(h_{SM} \rightarrow \tau\bar{\tau})
             + \Gamma_h^{(inv)} + \Gamma(h \rightarrow HH) , \\
  \Gamma_H &\simeq& \cos^2(\beta-\alpha)\Gamma(h_{SM} \rightarrow VV)
              + \left(\frac{\sin\alpha}{\sin\beta}\right)^2 \Gamma(h_{SM} \rightarrow q\bar{q}) \nonumber\\
          && + \left(\frac{\cos\alpha}{\cos\beta}\right)^2 \Gamma(h_{SM} \rightarrow \tau\bar{\tau})
             + \Gamma_H^{(inv)} ,
\end{eqnarray}
 with
\begin{eqnarray}
  \Gamma(h \rightarrow HH) &=& 
  \left( \sin\alpha \sin\beta (\cos2\alpha (-6 \lambda_1+3\lambda)-6 \lambda_1+\lambda) \right. \nonumber \\
&& \left. - \cos\alpha \cos\beta (\cos2\alpha (6 \lambda_2-3\lambda)-6 \lambda_2+\lambda) \right)^2
          \frac{v^2}{128\pi m_h}\sqrt{1-\frac{4 m_H^2}{m_h^2}} , \\
 \lambda &=& \lambda_3+\lambda_4+\lambda_5.
\end{eqnarray}          
          
\subsection{Dark matter annihilation cross section}
\begin{eqnarray}
 w(s) &\equiv &  \frac{1}{4}\int \overline{ |{\cal M}|}^2 d \rm{LIPS} , \\
 \overline{ |{\cal M}(\phi\phi\rightarrow b\bar{b})|}^2
   &=& 3 \left|\frac{(-\sigma_1 \sin\alpha\cos\beta+\sigma_2\cos\alpha\sin\beta)}{s-m_h^2+i m_h \Gamma_h} \frac{\cos\alpha}{\sin\beta}
 + \frac{(\sigma_1 \cos\alpha\cos\beta+\sigma_2\sin\alpha\sin\beta)}{s-m_H^2+i m_H \Gamma_H} \frac{\sin\alpha}{\sin\beta} \right|^2
 \nonumber \\ && \times  m_b^2 (s- 4m_b^2) , \\
 \overline{ |{\cal M}(\phi\phi\rightarrow \tau\bar{\tau})|}^2
   &=& \left|\frac{(-\sigma_1 \sin\alpha\cos\beta+\sigma_2\cos\alpha\sin\beta)}{s-m_h^2+i m_h \Gamma_h} \frac{\sin\alpha}{\cos\beta}
 - \frac{(\sigma_1 \cos\alpha\cos\beta+\sigma_2\sin\alpha\sin\beta)}{s-m_H^2+i m_H \Gamma_H} \frac{\cos\alpha}{\cos\beta} \right|^2 \nonumber \\ && \times  
 m_\tau^2 (s- 4m_{\tau}^2) .
\end{eqnarray}
%




\end{document}